\documentclass[11pt,a4paper]{article}
\usepackage{jheppub}
\usepackage{feynmp-auto}
\usepackage{tikz}
\usetikzlibrary{calc}
\usetikzlibrary{shapes,snakes}

\def\tr{\textrm{tr}}

\def\Im{\mathrm{Im\,}}
\def\sg(#1){\textrm{sign}(#1)}
\def\d{\mathrm{d}}
\def\bd {\boldsymbol}

\title{Infrared behaviour of the one-loop scattering equations and
  supergravity integrands}
\author{Eduardo Casali,}
\author{Piotr Tourkine}
\affiliation{Department of Applied Mathematics and Theoretical Physics,
Wilberforce Road, Cambridge CB3 0WA,
United Kingdom,}
\emailAdd{e.casali@damtp.cam.ac.uk}
\emailAdd{pt373@cam.ac.uk}

\abstract{The recently introduced ambitwistor string led to a
  striking proposal for one-loop maximal supergravity amplitudes,
  localised on the solutions of the ambitwistor one-loop scattering equations.
  However, these amplitudes have not been explicitly analysed, due to
  the apparent complexity of the equations that determine the
  localisation.
  In this paper we propose an analytic solution to the four-point
  one-loop scattering equations in the infrared (IR) regime of the
  amplitude.
  Using this solution, we compute the ambitwistor integrand and
  demonstrate that it correctly reproduces the four-graviton integrand
  in the IR regime. This solution qualitatively extends to $n$ points.
  To conclude, we explain that the ambitwistor one-loop scattering
  equations actually correspond to the standard Gross~\&~Mende
  saddle point.}

\preprint{DAMTP-2014-91}

\begin{document}
\maketitle

\section{Introduction}

Striking formulas to compute tree-level scattering amplitudes of spin
$0,1,2$ particles in arbitrary dimensions were proposed in a series of
papers by Cachazo, He and Yuan (CHY)
\cite{Cachazo:2013gna,Cachazo:2013hca,Cachazo:2013iaa}. In their
formalism, the amplitudes are obtained by localising certain integrands
on top of the solution set of a system of equations dubbed the ``scattering
equations''.
As shown in \cite{Mason:2013sva}, this formalism is properly
understood from first principles as arising from a chiral sigma model
called the ambitwistor string. This model consistently describes the
\textit{massless} sector of type II strings theories and reproduces
the CHY formulas.

Even more remarkable is that the formalism naturally yielded a
prescription to obtain loop amplitudes. In \cite{Adamo:2013tsa} a
one-loop $n$-graviton integrand for type II supergravity was
proposed. The structure of this amplitude is analogous to the
tree-level one; the loop level integrand is obtained by localising an
integral over the space of $n$-punctured worldsheets with one hole
(that is, tori), a loop momentum integral being unconstrained and left
to be done eventually. We review this prescription in
sec.~\ref{sec:review}.
The most immediate problem to solve is to find solutions to the
one-loop scattering equations. However, already at tree-level, finding
the solutions to the scattering equations is a hard task. The
equations are generically of degree $(n-3)!$, for which very little is
known analytically beyond six points
\cite{Kalousios:2013eca,Weinzierl:2014vwa,Dolan:2014ega}.
The one-loop scattering equations look even harder to solve. Firstly,
they inherit the complexity of the tree-level equations, since the
simple four-point one-loop case qualitatively corresponds to a
6-point tree after cutting open the loop. Moreover, they involve
elliptic functions instead of rational functions, and dealing with
these is technically challenging. Finally, the integrand itself is an
elliptic function, whose evaluation on the support of the equations
may seem \textit{a priori} only doable numerically.

In order to make progress we restrict ourselves to the loop-momentum
infrared (IR) kinematical region, which we introduce in
sec.~\ref{sec:infra-red-boundary}. We explain that the scattering
equations simplify but still contain non-trivial information. The main
results of the paper are then presented in sec.~\ref{sec:ir-sol-scat},
they can be summarised as follows:
\begin{itemize}
\item We show explicitly in sec.~\ref{sec:numer-struct-four} how
  the integrand of the four-graviton amplitude simplifies and reduces
  to the expected supersymmetrical kinematic factor. This holds for all
  kinematical regimes.
\item We solve analytically the scattering equations in the IR limit
  where three neighbouring propagators go on shell in
  sec.~\ref{sec:solving-se-triple}. This solution enables us to show
  in the following sec.~\ref{sec:computation-jacobian} that the
  ambitwistor integrand exactly reproduces the IR leading behaviour of
  the sum of scalar boxes of the field theory amplitude.
\item We also show that the
  four-point solution extends to $n$ points and that the $n$-point integrand
  qualitatively agrees with the expected IR divergence of the $n$-point amplitude.
\end{itemize}

Lastly, in sec.~\ref{sec:discussion}, we explain how the one-loop
ambitwistor saddle point actually coincides with the Gross \& Mende
saddle point; a connection which was solely understood at tree level
so far. More than a curiosity, this connection enables us to cross
check the consistency of technical details of our analysis, such as
the choice of a different bosonic propagator, and the absolute
normalisation of the ambitwistor loop momentum.
The last section \ref{sec:outlook} of the paper contains a review of
the literature on related works and a short outlook.

\section{Review of the one-loop ambitwistor string amplitude}
\label{sec:review}

\subsection{The amplitude}
\label{sec:amplitude}

The ambitwistor string models introduced in \cite{Mason:2013sva} are
worldsheet chiral conformal field theories (CFTs) which compute the
classical S-matrix of maximally supersymmetric gravity and gauge
theory. 
Their physical spectrum contains only these massless states, so there
is no need to take an limit $\alpha'\rightarrow 0$ as in conventional
string theory in order to decouple any sort of massive
modes.\footnote{In fact, there is no such thing as $\alpha'$ in these
  theories. The holomorphy of the ambitwistor string is more
  constraining than the smoothness conditions of ordinary string and
  rigidifies the worldsheet. From the CFT point of view, a consequence
  of this is that no $\langle XX\rangle$ type of contractions are
  allowed.}

However, the formalism is very close to the usual string theory one,
and one can write amplitudes on a worldsheet with holes, thereby
describing loop amplitudes, but involving these massless states
exclusively.
At genus one, the authors of \cite{Adamo:2013tsa} proposed an
expression for the one-loop amplitude for $n$ external particles as an
integral over the moduli space of a $n$-marked torus.
This integral is almost completely localised on the solutions to the
one-loop generalisation of the scattering equations; all the moduli --
the size of the torus and the position of the punctures -- are
fixed, but the loop-momentum integral remains.

For 10 dimensional supergravity, this integral has ultraviolet
divergences, which require introducing a cut-off. In this work we will
be rather cavalier about this and focus only on the \textit{integrand}
of the one-loop ambitwistor string amplitude. Also will we be not so
precise on the space-time dimensions, strictly speaking working in
$d=10$, but we will see along the text that the solution to the
scattering equations does not depend on which dimension we are working
in.\footnote{A four dimensional ambitwistor string construction with
  non-vanishing central charge was proposed in \cite{Geyer:2014fka}.
  It would be interesting to see what features of this model carry
  over to one loop.}

The genus one, $n$-graviton scattering amplitude in the ambitwistor
string receives two types of contributions, from the even and odd spin
structures of the torus, corresponding to physical CP sectors
of the amplitude. We denote the spin structures by bold greek indices $\bd
\alpha$, such that $\bd{ \alpha}=1$ is the odd one and
$\bd{\alpha}=2,3,4$ are the even ones.
The even spin-structure contribution is given\footnote{The odd spin
  structure does not contribute at four points, while our $n$ point
  considerations later in the text do not require us to write down the
  explicit form of the CP odd amplitude. It mostly contains a
  fermionic zero model integral in addition to the one present in
  (\ref{e:graveven}).} in terms of the the 10-dimensional
field $P^{\mu}$, $\mu=0,1,\cdots,9$, as follows
\begin{equation}
	\begin{aligned}
	\mathcal{M}_n^{1;\,{\rm even}}&= \delta^{10}\!\left(\sum_{i=1}^n k_i\right)\int 
	\d^{10}\ell\wedge\d\tau \prod_{j=2}^{n} \d z_{j} \
	    \bar\delta\!\left(P^2(z;\tau)\right)\,
	    \prod_{j=2}^n \bar\delta(k_j\cdot P(z_j)) \\
	&\hspace{4cm} \times\ \sum_{\bd{\alpha};\bd{\beta}=2,3,4}
	    (-1)^{\bd{ \alpha}+ \bd{\beta}} Z_{ \bd{\alpha}; \bd{\beta}}(\tau)
	\ {\rm Pf}(M_{ \bd{\alpha}})\,{\rm Pf}(\widetilde{M}_{ \bd{\beta}})\,.
	    \end{aligned}
\label{e:graveven}
\end{equation}

The above formula decomposes in three parts. 
The first line contains a measure and some delta functions. The
measure contains both a field-theoretic integration for the zeros
modes of the $P^{\mu}$ field, the 10-dimensionnal loop momentum
$\ell^{\mu}$, and a stringy worldsheet moduli integral. The delta
functions impose the scattering equations and localise the latter
intergation as a function of the former. The second line contains the
result of computing the CFT correlator between the vertex operators of
the external states.

The scattering equations and the associated Jacobian are universal for
massless scattering, we will see below that it contains information
about the scalar propagators of the field theory integrand.

The CFT correlator is written in this case as an even spin-structure
sum of the product of Pfaffians, dressed with partition functions
$Z_{\bd{\alpha},\bd{\beta}}$, see eq. \eqref{eq:Zdef}. It contains all
the information about the kinematics of the integrand; helicities and
momenta of the particles being scattered $k_{i}^{\mu},\epsilon^{\mu}$,
$i=1,\cdots,n$. The matrix $M_{\bd\alpha}$ is a generalisation of the
matrix in the CHY formula. It has the following form;
\begin{equation}
	M_{\bd\alpha} = \begin{pmatrix} 
			A & -{C}^{\rm T}\\
			C & B
		\end{pmatrix}\,,
\end{equation}
Its elements are 
\begin{equation}
	A_{ij} = k_i\cdot k_j\, S_{\bd\alpha}(z_{ij}|\tau) 	\qquad\qquad B_{ij} = \epsilon_i\cdot\epsilon_j \,S_{\bd\alpha}(z_{ij}|\tau)
	\qquad\qquad C_{ij} = \epsilon_i\cdot k_j\, S_{\bd\alpha}(z_{ij}|\tau)
\label{ABCg=1}
\end{equation}
and $A_{ii}= B_{ii}=0$. The diagonal entries of $C$ are 
\begin{equation}
	C_{ii} = \epsilon_i\cdot \ell\ \d z_i + \sum_{j\neq i} \epsilon_i\cdot k_j\, \partial G(z_{ij}|\tau)\d z_i \,,
\label{Cdiag}
\end{equation}
where we use the notation $z_{ij}=z_i-z_j$ and where
$\partial\equiv(\partial/\partial z)$ (respectively for $\bar\partial$). The function
$G(z_{ij}|\tau)$ is the bosonic propagator on the torus
\begin{equation}
 G(z|\tau)=
 -\ln\left|\frac{\theta_1(z|\tau)}{\partial\theta_1(0|\tau)}\right|^{2}+2\pi\frac{(\Im
   z)^2}{\Im\tau}\,.
\label{eq:Gdef}
\end{equation}
The functions
\begin{equation}
  S_{\bd \alpha}(z_{ij}|\tau) = \frac{\partial\theta_{1}(0;\tau)}{\theta_{1}(z_{ij};\tau)}\frac{\theta_{\bd \alpha}(z_{ij};\tau)}{\theta_{\bd \alpha}(0;\tau)}\sqrt{\d z_i}\sqrt{\d z_j}
\label{Szego}
\end{equation}
are the torus free fermion propagators, or Sz\"ego kernels, in the even
spin-structure $\bd{\alpha}$. The tilde matrix $\tilde M_{\bd\alpha}$
is defined in the same way as $M_{\bd\alpha}$ but with possibly
different polarisation vectors $\tilde\epsilon$. The
$Z_{\bd{\alpha; \beta}}$ are CFT partition functions in the
${\bd{\alpha}; \bd{\beta}}$ spin-structures
\begin{equation}
 Z_{\bd{\alpha;
     \beta}}=\frac{1}{\eta(\tau)^{16}}\frac{\theta_{\bd{\alpha}}(0|\tau)^4}{\eta(\tau)^4}\frac{\theta_{\bd{\beta}}(0|\tau)^4}{\eta(\tau)^4},
\label{eq:Zdef}
\end{equation}
They are defined in terms of the Dedekind eta function
\begin{equation}
  \label{eq:dedekind}
  \eta(\tau)=q^{1/24}\prod_{n=1}^{\infty}{(1-q^{n})}\,,
\end{equation}
and the Jacobi theta functions, themselves defined by Fourier-Jacobi $q$-expansions,
\begin{equation}
  \label{eq:qdef}
  q=e^{2i\pi\tau}\,,
\end{equation} 
as
\begin{equation}
  \label{eq:thetadef}
  \theta_{\bd{\alpha}}(z|\tau) = 
  \sum_{n\in\mathbb{Z}}
  q^{(1/2)(n-a/2)^{2}}e^{2i \pi (z-b/2)(n-a/2)}\,.
\end{equation}
Here $\bd{\alpha}:=(a,b)=(0,0),(0,1),(1,0)$ are the even characteristics and
$(1,1)$ is the odd one. In the $\bd{\alpha}=1,2,3,4$ notation used above, they
correspond to $\bd{\alpha}=3,4,2$ and $\bd{\alpha}=1$, respectively.

At this point, it might seem strange that \eqref{e:graveven} could ever
reproduce field theory amplitudes.
In the usual string theory, the contribution from the tower of massive
states is encoded by the infinite series expansion of the theta
functions, but in field theory there is no infinite tower of massive
states and we expect a rational integrand. 

\begin{figure}[t]
  \centering
  \includegraphics[scale=0.8]{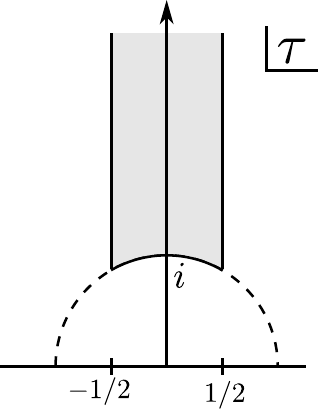}
  \caption{An $SL(2,\mathbb{Z})$ fundamental domain of the torus.}
  \label{fig:funddom}
\end{figure}

The way out comes from the scattering equations. Each solution
$\{z_i^*,\tau^*\}$ of the scattering equations should presumably be a
non-trivial, elliptic function of the external kinematics and loop
momentum.  If this expression is to reproduce the one-loop amplitude,
then it must be that once we evaluate the integrand on top of each
solution and sum over all of them, we eventually obtain a rational
function.
This is analogous to what occurs in the formulas for tree-level
scattering, where each solution involves very complicated algebraic
functions of the external kinematics. After summing over all the
solutions, the result is a rational function. We will see later that
this indeed occurs at one loop in a particular kinematical regime.

Before moving on, we note that in \cite{Adamo:2013tsa} the formula
\eqref{e:graveven} was shown to factorise as expected from a field
theory amplitude and that it is modular invariant provided the loop
momenta transforms in appropriate way. This means that the integration
region for the modular parameter $\tau$ is a fundamental domain in the
upper half plane (see fig.~\ref{fig:funddom}). Its only boundary is
located at $\Im\tau=\infty$ and physically corresponds to the infrared
(IR) regime of the amplitude, in virtue of the aforementioned
factorisation argument. This region is of crucial importance in this
work.

\subsection{The scattering equations}
\label{sec:rewriting-equations}

The most notable novel ingredient of these formulas is the
generalisation of the scattering equations to one loop. These are the
constraints imposed through the holomorphic delta functions in
\eqref{e:graveven}. As in the tree-level case, the one-loop scattering
equations relate the boundaries of the moduli space of curves to
factorisation channels of the amplitudes when one or more of the
Lorentz invariant kinematical factors approach zero. From the
worldsheet perspective, the geometric content of these equations is
that they enforce the vanishing of the quadratic differential
$P^2(z,\tau)=0$.

In \cite{Adamo:2013tsa}, the following representation for the
10-dimensional momentum $P^{\mu}$ field, was used;
\begin{equation}
 P^\mu(z)= \ell^\mu\d z+\sum_{i=1}^n k_i^\mu\partial G(z-z_i|\tau)\d z
\label{eq:Pmudef}
\end{equation}
where $G(z-z_i|\tau)$ is the bosonic propagator on the torus defined
in eq.~\eqref{eq:Gdef} and $\ell^\mu$ is a zero mode for $P^{\mu}$, in
other words it is the loop momentum.
$P^{\mu}$ is a meromorphic differential with at most simple poles,
thus $P^2$ is actually a {meromorphic} quadratic
differential. Since all the external momenta are on-shell, the double
poles of $P^2$ vanish. Therefore, one way to ensure that $P^2=0$ is to
require first that $n-1$ of its residues at its simple poles vanish
\begin{align}
\text{Res}_{z_i}(P^2(z))=2k_i\cdot \ell+2\sum_{j\neq i}k_i\cdot k_j\partial G(z_{ij}|\tau)=2k_i\cdot P(z_i)=0 
\label{eqres}
\end{align}
Since there is no meromorphic quadratic differential with only one
pole on the torus, $P^2$ is holomorphic on the support of these
equations. We can now enforce
\begin{align}
 P^2(z_0|\tau)=0
\label{eqhol}
\end{align}
at a some point on the torus. On the support of the first $n-1$
equations \eqref{eqres}, the last equation kills the holomorphic part
of the differential, which means that it is identically zero.  These
two sets of equations, \eqref{eqres} and \eqref{eqhol}, taken together,
are the scattering equations at one loop.

They can actually be written in numerous ways, depending
on how we choose to represent the bosonic propagator, and on how we
divide it into zero and nonzero modes. What constrains the possible
representations is that the field $P^{\mu}$ should obey the differential
equation
\begin{equation}
 \bar\partial P^\mu(z)=\sum_{i=1}^n k_i^\mu\bar\delta(z-z_i)\d z.
\end{equation}
which sets $P^{\mu}$ to to be a meromorphic differential on the torus
with residue $k_i^{\mu}$ at the pole $z_i$. In the following sections
we shall use a manifestly holomorphic version of the scattering
equations. This is a different choice than the representation of
\cite{Adamo:2013tsa} recalled in eq.~(\ref{eq:Pmudef}), where the aim
was to make the modular properties of the amplitude manifest. The
holomorphic choice simplifies the analysis considerably and it is
closer to the equations obtained by Gross and Mende
\cite{Gross:1987ar}, once the loop momentum is restored, as we discuss
in sec.~\ref{sec:discussion}.
The purely holomorphic version of the bosonic propagator,
$S_{1}$, is the Sz\"ego kernel in the spin structure $\bd{\alpha}=1$,
\begin{equation}
  \label{eq:szego}
  S_{1}(z|\tau)=\frac{\partial\theta_{1}(z|\tau)}{\theta_{1}(z|\tau)}\,,
\end{equation}
is related to the full propagator as
\begin{equation}
  \label{eq:S11toG}
\partial G=-S_{1}(z|\tau)-2i\pi \frac{\Im z}{\Im \tau}
\end{equation}
Hence, the relationship to the representation of \cite{Adamo:2013tsa} is
simple, we only need to redefine the loop momentum as
\begin{equation}
\ell^{\mu}\rightarrow \ell^{\mu}+ 2i\pi\sum_{i=1}^n k_{i}^{\mu}\frac{\Im(z-z_i)}{\Im(\tau)}.
\end{equation}
The local behaviours of both propagators are of course the same as on
the sphere
\begin{equation}
  \label{eq:local-szego}
  G(z|\tau),\,S_{1}(z|\tau)\underset {z\to 0}{\sim}\frac{1}{z}.
\end{equation}
For later use, we provide here the Fourier-Jacobi $q$-expansion of $S_{1}$
\begin{equation}
  \label{eq:FJszego}
  S_{1}(z|\tau)=\frac{\pi}{\tan(\pi z)} +4 \pi \sum_{n=1}^\infty
  \frac{q^{n}}{1-q^{n}} \sin(2n \pi z)\,.
\end{equation}
We finally introduce the following condensed notation\footnote{We
  trust the reader to not confuse the particle indices $i,j$ in the
  condensed notation with the spin structure indices. Unless explicitly
  stated the spin structure of the propagators is always odd.}
\begin{align}
 S_{ij} := S_{1}(z_{ij}|\tau).
\end{align}

The manifestly holomorphic scattering equations now read\footnote{Note
  that the $(n+1)$-th equation
  $ \ell\cdot k_1 +\sum_{j\neq 1}k_{1}\cdot k_{j}S_{1j}=0 $ holds
  automatically by momentum conservation}
\begin{subequations}
 \begin{align}
   &\ell\cdot k_i +\sum_{j\neq i}k_{i}\cdot k_{j}S_{ij}=0\,,\qquad i=2,\dots,n-1 \label{e:lk}\\
   &\ell^2+2 \sum_{i=1}^{n} \ell\cdot k_i\, S_{0i} +
   \sum_{i\neq j}^{n} k_i\cdot k_j\,
   S_{0i}S_{0j} =0\,.
   \label{e:l2l}
 \end{align}
\end{subequations}
We can use the equations \eqref{e:lk} to write equation \eqref{e:l2l} as
\begin{equation}
  0=\ell^2-2\sum_{1\leq i< j\leq4} 
  k_i\cdot k_j
  \left(S_{0i}S_{ij} +S_{j0}S_{0i}+S_{ij}S_{j0}
\right)\,.
  \label{e:l2SEsym}
\end{equation}
It is now easy to check that this equation has no pole in $z_0$; since
it is a holomorphic elliptic function on $z_0$ without any poles, by
Liouville theorem it has to be a constant.

\subsection{The Jacobian}
\label{sec:jacobian}

The universal contribution from the scattering equations to the
integrand comes from the Jacobian that appears when solving the delta
function constraints. This Jacobian has to contain all the information
of the scalar propagators of the amplitude, as it does in the CHY
formulas, except that at one-loop there is an extra loop momentum
integral that is not localised. The structure of the Jacobian is (we
denote $z$ derivatives with $'$)
\begin{align}\label{Jacobian}
 J=\left(\begin{array}{c|c}
    A_{ij} & B_i \\ \hline
    C_j & D
   \end{array}\right)
\end{align}
where 
\begin{equation}\label{Amatrix}
 A_{ij} = \begin{cases}\displaystyle
          k_i\cdot k_j S'_{ij},&\text{if }i\neq j\,,\\
          \sum_l k_l\cdot k_i S'_{il},&\text{if }i=j\,,\\
         \end{cases} 
\end{equation}
and
\begin{align}
B_i&= \ell\cdot k_i S'_{0i} +\sum_j k_i\cdot k_j S_{j0}S'_{i0}\,,\\ \label{Bcolumn}
C_i&= \sum_j k_i\cdot k_j \partial_\tau S_{ij}\,,\\ \label{Cline}
D&=\sum_i \ell\cdot k_i\partial_\tau S_{i0} + \sum_{j\neq i}k_i\cdot k_j S_{i0}\partial_\tau S_{j0}\,.
\end{align}

After solving the scattering equations, the integrand for the
amplitudes is computed by evaluating the Pfaffians and the Jacobian on
these solutions and summing over all of them. Schematically, this writes
\begin{equation}
  \label{eq:claim}
  \sum_{\mathrm{solutions}}\frac{\mathrm{Pf}(M)\mathrm{Pf}(\tilde M)}{J} =
  ``\mathrm{generalized\ integrand}"\,,
\end{equation}
where the right hand side stands for the result of bringing under the
same integral symbol the field theory integrands corresponding to
the the various Feynman graphs.

\section{Infrared Behaviour at One Loop}
\label{sec:infra-red-boundary}

In this section, we wish to describe both in the ambitwistor and in
field theory the IR regime in which we will explicitly solve the
scattering equations in sec.~\ref{sec:ir-sol-scat}. We start by
recalling the geometry associated to the IR pinching limit in the
ambitwistor string, then we briefly discuss the resulting tree-level
forward scattering in the CHY formalism. We finally introduce the
triple pinching limit which interests us. Throughout the rest of the
paper, we use the four point Mandelstam kinematic invariants defined
by $s=(k_1+k_2)^2,\,t=(k_1+k_4)^2,\,u=(k_1+k_3)^2$.

\subsection{Boundary behaviour of the ambitwistor amplitude}
\label{sec:bound-behav}

Solving the scattering equations at one loop for general kinematics is
a daunting task. Here we study them in the IR regime of the
amplitude, where the equations simplify and, guided by simple
numerics, we are able to find analytical solutions. The factorisation
of the amplitude at the boundaries of the modular space was already
studied in \cite{Adamo:2013tsa} and follows the general structure of
\cite{Adamo:2013tca}. 

Let us describe first some elements of this pinching limit. Let us
consider a kinematic regime in which $\ell^{2}\to0$. We wish to see
that the ambitwistor integrand, more precisely the Jacobian, produces
a $1/\ell^{2}$ term. 

It was demonstrated in \cite{Adamo:2013tsa} that in
this limit, the parameter $q$ defined in eq.~\eqref{eq:qdef}
can be consistently considered to vanish as well, $q\rightarrow 0$,
for certain solutions of the scattering equations. The converse is not
necessarily true; in principle, there could be solutions for which
$\ell^2\rightarrow 0$ but $q$ stays finite and our analysis won't be
sensitive to those solutions. By general worldsheet factorisation
arguments we believe that even if such solutions exist it do not
contribute to IR divergences. 

At $q=0$ and $\ell^{2}=0$, the $n$ scattering equations reduce to the
following $n-1$ ones
\begin{equation}
  \label{eq:2}
  P\cdot k_i(z_i)=\ell\cdot k_i+\sum_{j\neq i}
  \frac{\pi k_i\cdot k_j}{\tan(\pi z_{ij})} = 0.
\end{equation}
where we kept only the first term of the propagator in the expansion
eq.~(\ref{eq:FJszego}).  The last equation $P^{2}=0$ of
eq.~\eqref{e:l2l} is automatically satisfied at $q=0$; the
finite piece cancels due to the a trigonometric identity, somewhat
analogous to a partial fraction decomposition
\begin{equation}
  \label{eq:5}
  \frac{1}{\tan(\pi z_{ij})\tan(\pi z_{jk})}+  \frac{1}{\tan(\pi
    z_{jk})\tan(\pi z_{ki})}+  \frac{1}{\tan(\pi z_{ki})\tan(\pi z_{ij})}=-1\,,
\end{equation}
valid for any set of three complex numbers $z_{i},z_{j},z_{k}$.

At this stage, the choice of which propagator to use is immaterial
since $q=0$ is equivalent to $1/\Im \tau=0$, so that both propagators
coincide. We will argue now that using the full propagator obscures
the correct $1/\ell^{2}$ behaviour, thereby motivating our choice of a
holomorphic representation.

Consider the case of a large but not infinite $\Im \tau$, or small but
nonzero $q$. If we work with the full propagator of
eq.~(\ref{eq:Gdef}), i.e. the one with the non-holomorphic term, the
$\epsilon=1/\Im \tau$ correction is much bigger than corrections of
order $q$, so it makes sense to consider corrections of order
$\epsilon$, such that $z_{i}=z_{i}^{0}+\epsilon z_{i}^{\epsilon}$ is a
new solution to the scattering equations.

The first $P(z_{i})\cdot k_{i}$, $i=1,\cdots n-1$ equations are
still satisfied at order zero while the $O(\epsilon)$ terms give a
system of linear equations for the $z_{i}^{\epsilon}$.
Once plugged back in the last equation $P^{2}(z_{0})=0$, the zeroth
order cancels again but the $O(\epsilon)$ seems to undergo no further
obvious cancellations, indicating that $\epsilon$ is of the order of
the zero mode part $\ell^{2}$. This, \textit{a priori}, is a
possibility. Knowing that we expect the leading infrared behaviour of
our integrand to be $1/\ell^{2}$, it means that we want our Jacobian
to be of order $1/\epsilon$, that is, $\Im \tau$. As the analysis
below will demonstrate, the presence of $\tau$ derivatives in the
Jacobian always produces order $O(\epsilon^2)$ terms due to the fact that
$\partial_{\tau}(1/\Im \tau)=(2i)^{-1} (\Im\tau)^{-2}$. This second
order contributions to the Jacobian in turn seemingly gives an
incorrect IR behaviour of the form
$\dfrac{\d \ell}{\epsilon^2}\sim \dfrac{\d \ell}{\ell^4}$ instead of
the $1/\ell^2$ expected.

On the other hand, if we drop the non-holomorphic part of the
propagator, the first small correction to be turned on is of order
$q$.  The same analysis as above holds, but with
$\epsilon=q\sim \ell^2$. This is easily seen to produce the correct
qualitative IR behaviour since the $\tau$ derivatives do not change
anymore the overall degree of $\epsilon$;
$\partial_\tau q = 2 i\pi q$. This motivates our choice to adopt
purely holomorphic propagators from then on. We will come back to this
point when we discuss the connection with the Gross-Mende saddle
point.

Let us examine the behaviour of the Jacobian \eqref{Jacobian} on the
support of solutions for which $\ell^{2}\to0$ implies
$q\rightarrow 0$.  The propagators themselves reduce to a $1/\tan$
trigonometric function, as we saw in eq.~(\ref{eq:2}). The derivatives
of the propagator with respect to the coordinates $z_{i}$ (denoted
$S'$) are finite,
\begin{equation}
S'_{ij}\rightarrow -\frac{\pi^{2}}{\sin^2(\pi z_{ij})}+O(q)\,,
\end{equation}
but the $\tau$ derivatives are of order $q$:
\begin{equation}
\partial_\tau S_{ij}=8 i \pi^{2} q\sin(2\pi z_{ij})+O(q^{2})\,.
\end{equation}
Therefore, the last line of the Jacobian \eqref{Jacobian} is
proportional to $q$, which means that $|J|\rightarrow q|M|$ where $M$
has no other dependence on $q$ at leading order. Since
$\ell^2\propto q $ for small $q$ this explains why the Jacobian does
produce generically the scalar propagator that is going on shell, schematically;
\begin{equation}
  \label{eq:schemjacq}
  \frac{1}{\mathrm{Jacobian}}\propto \frac{1}{\ell^{2}}\,.
\end{equation}
We will see soon an explicit implementation of this with three adjacent
propagators going on-shell.

\begin{figure}[t]
  \centering
  \includegraphics[scale=0.8]{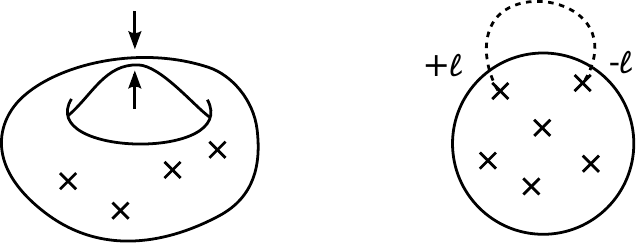}
  \caption{4-point pinched torus creates a 6-point sphere with two
    back-to-back momenta.}
  \label{fig:toruspinch}
\end{figure}

We end this section by recalling the geometry resulting from this
pinching limit. As explained in \cite{Adamo:2013tsa}, the
factorisation properties of the ambitwistor worldsheet in the $q\to0$
limit are very reminiscent of the traditional picture of string
theory. In particular, the fact that the torus pinches in the limit is
completely compatible with factorisation of the amplitude on the
$\ell^{2}=0$ channel. What is left can be interpreted as the forward
limit of an $(n+2)$-point tree-level amplitude, where the two new
punctures have back to back momentum $\ell^{\mu}$ and $-\ell^{\mu}$,
see figure \ref{fig:toruspinch}. Since the external kinematics are not
generic the number of independent solutions is smaller in this limit.

Numerically (using the simple \texttt{NSolve} routine of
\texttt{Mathematica}), we find, at 6,7 and 8 points, 2,12 and 72
solutions respectively. A reasonable conjecture for the generic
pattern of the number of solutions is $(n-3)!-2(n-4)!$;
\begin{equation}
  \label{eq:nsolsforward}
  N_{\mathrm{sols}}^{\mathrm{forward-tree}} = (n-3)!-2(n-4)!\,.
\end{equation}
We have no satisfactory proof of this, but it would
be very interesting to have one, maybe in the lines the recursive soft
limit used in \cite{Cachazo:2013gna}. In table~\ref{tab:PolesvsN}, we
display the known number of solutions for generic kinematics, the
number of solutions in the forward limit at low number of points and
the number of trivalent diagrams at $n$ points. This emphasises that
the number of solutions is much smaller than the number of diagrams at
tree level.

\begin{table}[hb]
  \centering
  \begin{tabular}{|l|c|c|c|}
    \hline
    $n$ & $N_{\mathrm{sols}}^{\mathrm{tree}}$ &
                                $N_{\mathrm{sols}}^{\mathrm{forward-tree}}$&
                                                                             Number of
                                                                             cubic
                                                                             graphs\\
    \hline 
    4&1&$\emptyset$&3\\
    5&2&$\emptyset$&15\\
    6&6&2&105\\
    7&24&12&945\\
    8&120&72&10395\\
    \hline
  \end{tabular}
  \caption{Number of solutions to the tree-level scattering equations (known to
    be $(n-3)!$), number of solutions in the forward kinematics, number
  of cubic graphs; $(2n-5)!!$.}
  \label{tab:PolesvsN}
\end{table}

We therefore expect that the number of boundary solutions to the
one-loop scattering equations is equal to the number of solutions in
the tree level forward kinematics, making it equal to
$(n-1)!-2(n-2)!$. We observed numerical agreement with this claim at 4
and 5 points, while we did not try to solve numerically the one-loop system
at 5 points with a non-vanishing $q$. We already mentioned that if there
exists additional solutions which are not sent to the boundary of the
moduli space, our analysis is insensitive to it (as we want to
capture only the IR divergences), therefore the total number of
solutions is at any rate bounded by the number of conjectured
tree-level forward solutions;
\begin{equation}
  \label{eq:nsols}
  N_{\mathrm{sols}}^{1-\mathrm{loop}}\geq (n-1)!-2(n-2)!\,.
\end{equation}

\subsection{Three propagators on-shell}
\label{sec:triple-cut-like}

\begin{figure}[t]
  \centering
\begin{fmffile}{IR}
\fmfframe(0,10)(10,0){ \begin{fmfgraph*}(100,50)
	\fmflabel{$i$}{gi}
	\fmflabel{$j$}{gj}
	\fmfleft{gi,g1,g0}
	\fmfright{gj,g4,g5}
	\fmf{dashes}{g0,v01,g1}
	\fmf{plain}{gi,vi}
	\fmf{plain}{gj,vj}
	\fmf{dashes}{g4,v45,g5}
	\fmf{fermion,label=$\ell-k_{j}$,label.dist=2pt}{v45,vj}
	\fmf{fermion,label=$\ell$,label.dist=-14pt}{vj,vi}
	\fmf{fermion,label=$\ell+k_{i}$}{vi,v01}
\end{fmfgraph*} }
\end{fmffile}
\caption{Typical IR divergences in theories of gravity.}
\label{fig:IRdiv}
\end{figure}
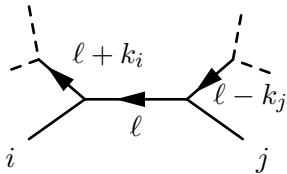

The kinematic regime in which we will be able to produce analytic
results is characterised by the fact that three adjacent propagators
are going on shell, $\ell^{2},(\ell+k_{i})^{2},(\ell-k_{j})^{2}\to0$.
From the point of view of the pinched worldsheet described before,
this can be seen as a sort of a double collinear limit, where we tune
the loop momentum $\ell^{\mu}$ becomes collinear with two external
particles $k_i^{\mu}$ and $k_j^{\mu}$.  The leading infrared
divergence originates from the configuration where the legs $i$ and
$j$ are adjacent, as pictured in fig.~\ref{fig:IRdiv},
which results in the following behaviour
\begin{equation}
  \label{eq:deepir2}
  \text{leading~IR} \sim \frac{1}{(\ell\cdot k_{i} )\ell^{2} (\ell\cdot k_{j})}
\end{equation}
up to an overall product of propagators corresponding to the ordering
of the graph. In gauge theory, these would be dressed with appropriate
colour factors selecting possible divergences. In gravity or QED
\cite{Badger:2008rn} this is not the case, since all orderings
contribute equally, therefore we can regroup these diverging terms
under the same integration.

\begin{figure}[h]
\centering
    \parbox{80pt}{ \begin{fmffile}{box1234}
        \fmfframe(30,10)(30,0){ \begin{fmfgraph*}(60,48)
            \fmflabel{$1$}{g1} \fmflabel{$2$}{g2} \fmflabel{$3$}{g3}
            \fmflabel{$4$}{g4} \fmfleft{g1,g2} \fmfright{g4,g3}
            \fmf{plain}{g1,v1} \fmf{plain}{g2,v2} \fmf{plain}{g3,v3}
            \fmf{plain}{g4,v4}
\fmf{fermion,label=$\ell-k_2$,label.side=left,tension=0.4}{v1,v2}
\fmf{fermion,label=$\ell$,label.side=left,tension=0.4}{v2,v3}
\fmf{fermion,label=$\ell+k_3$,label.side=left,tension=0.4}{v3,v4}
            \fmf{plain,tension=0.4}{v4,v1}
        \end{fmfgraph*} }
    \end{fmffile} } 
    \parbox{80pt}{
\begin{fmffile}{box4231}
  \fmfframe(30,10)(30,0){ \begin{fmfgraph*}(60,48) \fmflabel{$4$}{g1}
      \fmflabel{$2$}{g2} \fmflabel{$3$}{g3} \fmflabel{$1$}{g4}
      \fmfleft{g1,g2} \fmfright{g4,g3} \fmf{plain}{g1,v1}
      \fmf{plain}{g2,v2} \fmf{plain}{g3,v3} \fmf{plain}{g4,v4}
\fmf{plain,tension=0.4}{v1,v2}
\fmf{fermion,label=$\ell$,label.side=left,tension=0.4}{v2,v3}
\fmf{plain,tension=0.4}{v3,v4}
\fmf{plain,tension=0.4}{v4,v1}
\end{fmfgraph*} }
\end{fmffile} }
\parbox{80pt}{\begin{fmffile}{box1324}
\fmfframe(30,10)(30,15){ \begin{fmfgraph*}(60,48)
	\fmflabel{$1$}{g1}
	\fmflabel{$3$}{g2}
	\fmflabel{$2$}{g3}
	\fmflabel{$4$}{g4}
	\fmfleft{g1,g2}
	\fmfright{g4,g3}
	\fmf{plain}{g1,v1}
	\fmf{plain}{g2,v2}
	\fmf{plain}{g3,v3}
	\fmf{plain}{g4,v4}	 
        \fmf{plain,tension=0.4}{v1,v2}      \fmf{fermion,label=$\ell$,label.side=left,tension=0.4}{v2,v3}
\fmf{plain,tension=0.4}{v3,v4}
\fmf{plain,tension=0.4}{v4,v1}
\end{fmfgraph*} }
\end{fmffile}}
\parbox{80pt}{\begin{fmffile}{box4321}
\fmfframe(30,10)(30,0){ \begin{fmfgraph*}(60,48)
	\fmflabel{$4$}{g1}
	\fmflabel{$3$}{g2}
	\fmflabel{$2$}{g3}
	\fmflabel{$1$}{g4}
	\fmfleft{g1,g2}
	\fmfright{g4,g3}
	\fmf{plain}{g1,v1}
	\fmf{plain}{g2,v2}
	\fmf{plain}{g3,v3}
	\fmf{plain}{g4,v4}	 
\fmf{plain,tension=0.4}{v1,v2}      \fmf{fermion,label=$\ell$,label.side=left,tension=0.4}{v2,v3}
\fmf{plain,tension=0.4}{v3,v4}
\fmf{plain,tension=0.4}{v4,v1}
\end{fmfgraph*} }
\end{fmffile}
}\\
\hspace{50pt} 
a)
\hspace{70pt} 
b) 
\hspace{70pt} 
c) 
\hspace{70pt} 
d)
\hspace{50pt} 
\caption{The four boxes that contribute to the IR divergence}
\label{fig:boxes}
\end{figure}
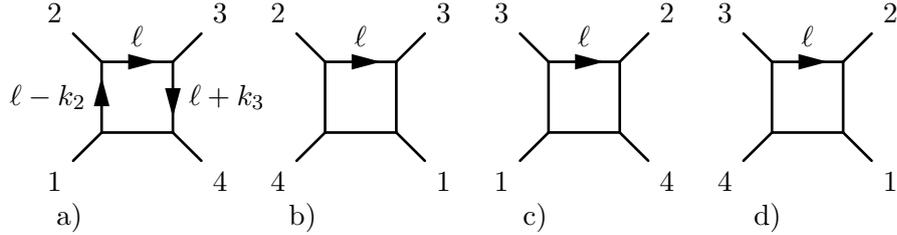

As we will demonstrate in the next section, the explicit solution of
the scattering equations in this IR regime will modify the scaling of
$q$ to
\begin{equation}
  q\propto \ell^2(\ell\cdot k_i)(\ell\cdot k_j)\,.
\end{equation}
The qualitative IR behaviour of the ambitwistor Jacobian then follows
from the fact that the Jacobian is of order $q$ in this limit and the
leftover determinant is finite and nonzero, as in
eq.~(\ref{eq:schemjacq}). At four points, this can be made very
precise.  Consider taking $\ell^2\rightarrow0$ as well as taking the
loop momenta to be collinear with particles 2 and 3. The boxes which
contribute to the the leading IR divergence are given in figure
\ref{fig:boxes}.

Their contribution is
\begin{equation}
  \begin{array}{cc}
    \mathrm{box}_a=\dfrac{1}{2\ell\cdot k_4+s}\;\;\; & \mathrm{box}_b=\dfrac{1}{-2\ell\cdot k_4+u}\\
    \mathrm{box}_c=\dfrac{1}{2\ell\cdot k_4+u}\;\;\; &  \mathrm{box}_d=\dfrac{1}{-2\ell\cdot k_4+s}\\
  \end{array}
\end{equation}
up to a global diverging factor of
\begin{equation}
  \label{eq:IRdiv}
   \frac{-1}{4 (\ell\cdot k_{2} )\ell^{2} (\ell\cdot k_{3})}\,.
\end{equation}
Bringing all these diverging integrands under the same integral symbol,
we obtain the leading IR divergence
\begin{equation}
  \label{eq:totIR}
    \frac{-1}{2 (\ell\cdot k_{2} )\ell^{2} (\ell\cdot k_{3})}
\left(\frac{-stu + t(2\ell\cdot k_4)^2}{(s^2-(4\ell\cdot k_4)^2)(u^2-(4\ell\cdot k_4)^2)}\right)\,.
\end{equation}
It is this non-trivial factor, including its functional dependence on
the last propagator $\ell\cdot k_{4}$, that we will demonstrate to arise from
the ambitwistor integrand in the following section.

\section{IR solution}
\label{sec:ir-sol-scat}

The object that we want to compute is composed of a numerator, the
Pfaffians, and a denominator, the Jacobian. These are evaluated on top
of the solutions of the scattering equations and then summed over them
all.

In this section, we compute first the Pfaffian for the four graviton
amplitude. We observe that target space supersymmetry factors out of
the integral all the kinematical dependence of the numerator. Then we
solve the scattering equations, and plug these solutions back into the
Jacobian.  The \texttt{Mathematica} evaluation of the $4\times4$
determinant on the support of the solution simplifies to a
single term, precisely the one needed to match the field theory
integrand.

\subsection{Numerator structure}
\label{sec:numer-struct-four}

The Pfaffians entering \eqref{e:graveven} may seem to be extremely
complicated objects, as they depend on various theta functions and
derivatives thereof.  It is far from obvious that these objects not
only give rational functions of the kinematic invariants but also
reproduce the very simple integrands of maximal supergravity. These
type of spin structure sums are however well known in RNS string
amplitudes, for which simplifications are known to arise due Riemann's
theta-function identities (see for example
\cite{mumford1983tata}). The one we need here is
\begin{equation}
  \label{e:Jacobi}
  \sum_{\alpha=1,2,3,4} (-1)^{\alpha-1}
  \prod_{i=1}^4 \theta_\alpha(v_i)
  =-2\prod^4_{i=1} \theta_1(v'_i)\,,
\end{equation}
with $v'_1={1\over2} (-v_1+v_2+v_3+v_4)$,
$v'_2={1\over2}(v_1-v_2+v_3+v_4)$, $v'_3={1\over2}(v_1+v_2-v_3+v_4)$,
$v'_4={1\over2}(v_1+v_2+v_3-v_4)$.
This identity gives rise to four vanishing identities 
\begin{equation}
   \begin{aligned}
       &\sum_{\alpha=2,3,4} (-1)^{\alpha-1} 
       {\theta_\alpha(0|\tau)^4
       \over \eta(\tau)^{12}} (\tau) =0
       \,,\\
       &\sum_{\alpha=2,3,4} (-1)^{\alpha-1} 
       {\theta_\alpha(0|\tau)^4 \over \eta(\tau)^{12}} 
       \prod_{r=1}^n S_\alpha(z_r)=0\,,
   \end{aligned}
\label{e:Riemann03}
\end{equation}
for $n=1,2,3$ and the $z_r$'s arbitrary. The first non-vanishing identity is
\begin{equation}
  \sum_{\alpha=2,3,4} (-1)^{\alpha-1} 
  {\theta_\alpha(0|\tau)^4 \over \eta(\tau)^{12}} 
  \prod_{i=1}^4 S_\alpha(z_i|\tau)=-(2\pi)^4\,,
  \label{e:Riemann4}
\end{equation}
for $z_1+\cdots+z_4=0$. In order to write \eqref{e:Riemann4}, we used that 
\begin{equation}
\partial_z
\theta_1(0|\tau)=\pi\theta_2(0|\tau) \theta_3 (0|\tau)\theta_4 (0|\tau)=2\pi\eta^3(\tau)\,,
\end{equation}
where we introduced Dedekind $\eta$ functions
in order to have the partition functions $Z_{\bd{\alpha}}$ defined in
eq.~(\ref{eq:Zdef}) explicit in the
left hand side of eqs.~(\ref{e:Riemann03}),(\ref{e:Riemann4}).

The consequence of these identities, as in string theory, is that the
0, 1, 2 and 3-point amplitudes vanish by supersymmetry. This does not
exclude the possibility that scattering equations, possibly deformed,
may be devined for these cases, but this question out of the scope of
this work. The 4-point one simplifies
considerably and the whole ambitwistor numerator boils down to a
single kinematical term, the
$t_{8}F^{4} t_{8}\tilde{F}^{4}=t_{8}t_{8}R^{4}$ tensor. This is the
only kinematic invariant at four points allowed by maximal
supersymmetry, of the form $R^{4}$.\footnote{The field strength $F^{\mu\nu}$ is the
  linearized field strength defined by
  $F^{\mu\nu} = \varepsilon^{\mu} k^{\nu}-k^{\mu}\varepsilon^{\nu}$
  and $R^{\mu\nu\rho\sigma} = F^{\mu\nu}F^{\rho\sigma}$. Then $t_8$
  tensor is defined in \cite[Appendix~9.A]{Green:1987sp}, it is given
  by
  $t_8F^4=4\tr(F^{(1)}F^{(2)}F^{(3)}F^{(4)})-
  \tr(F^{(1)}F^{(2)})\tr(F^{(3)}F^{(4)})+{\rm perms}\,(2,3,4)$,
  where the traces are taken over the Lorentz indices. In the
  spinor-helicity formalism one has
  $2t_8 F^4 = \langle12\rangle^2 [34]^2 $ and
  $4t_8t_8 R^4 =\langle12\rangle^4 [34]^4 $. Note also that
  $\langle12\rangle^2 [34]^2 = i st A^{tree}(1^-,2^-,3^+,4^+)$ where
  $A^{tree}$ is the tree level four graviton amplitude.}  The four
point amplitude is given by the simple integral
\begin{equation}
I_4=t_{8}t_{8} R^{4}\int \d\tau\d z_2\d z_3\d
z_4\bar\delta(P^2(z_0))\bar\delta(k_2\cdot P(z_2))\bar\delta(k_3\cdot
P(z_3))\bar\delta(k_4\cdot P(z_4))\,,
\label{eq:I4}
\end{equation}
Thus, the leftover physics of the integrand is captured solely by the
Jacobian. Its evaluation on top of the solutions of the scattering
equations should reproduce the the one-loop four-graviton integrand,
which is a simple sum of scalar box integrands \cite{Green:1982sw}.

This also gives a tempting interpretation of integrals of the type of
$I_{4}$ for generic $n$ as a representation of scalar $n$-gons
integrals.

\subsection{IR solution to the four-point one-loop scattering equations}
\label{sec:solving-se-triple}

For definiteness, let us define the kinematical IR regime by fixing
$\ell \cdot k_1$ and $\ell \cdot k_4$ and sending $\ell \cdot k_2\to0$
and $\ell \cdot k_3\to0$, with $\ell \cdot k_2< \ell\cdot k_3$.
We start the analysis by summarizing the results of a numerical study
that we performed for $q=0$ and $\ell^{2}=0$ and variations to small
but nonzero values. As we discussed above, in this regime, the
equations that we need to solve are similar to the 6-point tree-level
equations, which are easy to solve numerically.

The first outcome of the numerics is that there are only two solutions,
complex conjugate to one another. We checked that this still holds after
perturbing the system and finding for solutions with small $q$ .

The second one, notably important, is that the leading part of the
positions of the vertex operators scale as;
\begin{equation}
  \begin{aligned}
    i \pi z_{2}&= \log(\sqrt{\ell \cdot k_{2} c_{2}})\\
    i \pi z_{3}&= -\log(\sqrt{\ell \cdot k_{3}c_{3}})
  \end{aligned}
\label{eq:scalingpkpk23}  
\end{equation}
where $c_2$ and $c_3$ are complex constants of mass dimension $(-2)$,
to be determined.  Finally, it should be noted that the signs are
obtained for a given kinematic configuration, for which in particular
$\ell \cdot k_2< \ell\cdot k_3$. For consistency, in other kinematical
configurations the signs might change.

We can now declare that \eqref{eq:scalingpkpk23} is our
\textit{ansatz}, in which $c_{2}$, $c_{3}$ and $z_{4}$, or rather
\begin{equation}
  \label{eq:c4def}
  c_4:=\exp(-2 i \pi z_4)\,,
\end{equation}
are unknowns to be determined to first order in
$q, \ell\cdot k_{2},\ell\cdot k_{3}$. In that manner, the scattering
equations can be simplified by Taylor expanding the propagators
\begin{equation}
  \label{eq:identities}
\begin{aligned}
  i\cot(\pi z_{21})&=1+2\ell\cdot k_2 c_2\\
  i\cot(\pi z_{23})&=1+2\ell\cdot k_2 \ell\cdot k_3 c_2 c_3\\
  i\cot(\pi z_{24})&=1+2\ell\cdot c_2 c_4\\
  -i\cot(\pi z_{31})&=1+2\ell\cdot k_3 c_3\\
  -i\cot(\pi z_{34})&=1+2\ell\cdot k_3 c_3/c_4
\end{aligned}
\end{equation}
where we omitted the mention $+O(q)$ for clarity on the right hand
side of these equations.  It is easy to derive similar rules for
any trigonometric function of the same arguments, so we shall not
display them here. They are nonetheless important for the explicit
evaluation of the Jacobian.

With these, the $P\cdot k_{4}$ scattering equation simplifies
drastically and one obtains immediately
\begin{equation}
  \label{eq:cotz4}
  \pi \cot(\pi z_4)= \frac{\ell\cdot k_4}{k_1\cdot k_4}+i \pi \frac{s-u}t
\end{equation}
from which we extract $c_{4}$.

The scattering equations for $P\cdot k_{2}$ and  $P\cdot k_{3}$ are rewritten at
leading order 
\begin{equation}
  \label{eq:sceqns23IR}
\begin{aligned}
  2\ell\cdot k_2 - i s (1+2\ell \cdot k_2 c_2)
- i t (1+2\ell \cdot k_2\ell \cdot k_3 c_2c_3)
- i u (1+2\ell \cdot k_2 c_2c_4)&=0\,,\\
  2\ell\cdot k_3 + i u (1+2\ell \cdot k_3 c_3)
+ i t  (1+2\ell \cdot k_2\ell \cdot k_3 c_2c_3)
+ i s (1+2\ell \cdot k_3 c_3/c_4)&=0\,.
\end{aligned}
\end{equation}
After using momentum conservation, these reduce to a degenerate
system of quadratic equations whose unique solution is given by
\begin{equation}
  \label{eq:c2c3}
\begin{aligned}
  c_2&=\frac{i\ell\cdot k_4-\pi u}{\pi t\ell\cdot k_4} \,,\\
  c_3&=-\frac{i\ell\cdot k_4+\pi s}{\pi t\ell\cdot k_4}\,,\\
  c_4&=-\frac{\pi s+ i\ell\cdot k_4}{\pi u-i\ell\cdot k_4}\,,
\end{aligned}
\end{equation}
where we displayed the value of $c_{4}$. Of course we checked
numerically the agreement of this solution with the numerical data.

At this point, we turn back to the $P^{2}(z_{0})$ scattering equation
which determines $q$ to first order. We need to consider the new
scaling \eqref{eq:scalingpkpk23} in this limit limit. Using the
Fourier-Jacobi expansion \eqref{eq:FJszego}, we see that the
coefficients of $q$ include sinefunctions. These produce diverging
terms when its arguments involve the positions of the vertex operators
which become collinear to $\ell^\mu$. In particular, it is not hard to
see in \eqref{e:l2SEsym} that the most diverging term will come from
$\sin(2\pi z_{23})$, so that
\begin{equation}
  \label{eq:qpk2pk3}
  0=\ell^2+4 \pi^2 q k_2\cdot k_3 \left(S_{23}S_{30}+S_{32}S_{20}\right)\big|_{(q)}\,,
\end{equation}
at leading order.
To extract the exact value of this term, we use the independence of
$P^{2}(z_{0})$ with respect to $z_{0}$ and set $z_0$ to $1/2$. In this
case, the $\cot(\pi z_{20})$ and $\cot(\pi z_{30})$ terms just become
$\tan$'s which are readily evaluated to $\pm i$, as in
\eqref{eq:identities} (recall $z_{1}=0$). In total we are left with
\begin{equation}
  \label{eq:scalqpk2pk3}
  q=-\frac{c_2 c_3}{8\pi^2 k_2\cdot k_3}\ell^2(\ell\cdot k_2)(\ell \cdot k_3)\,.
\end{equation} 
This equation indicates that the scaling of $q$ is
not only dictated by the $\ell^2\to0$ but also by the collinear
$\ell \cdot k_{2}\to0$ and $\ell \cdot k_{3}\to0$ and other kinematic
invariants, as claimed in sec.~\ref{sec:triple-cut-like}.

The objective of the following computation is to evaluate the
Jacobian, and verify that it creates no further divergence that would
change this IR behaviour, and match it the field theory result
\eqref{eq:totIR}.

\subsection{Computation of the Jacobian}
\label{sec:computation-jacobian}

We observe first that, $q$ being stripped off the Jacobian, no more
factors of $\ell\cdot k_{2}$ or $\ell\cdot k_{3}$ contribute at first
order\footnote{There is a possible divergent piece in the Jacobian. It
  is not hard to see that it terms of order
  $\ell\cdot k_2\ell\cdot k_3$ always make these terms individually
  finite. This pattern extends to higher points.}, making this
stripped determinant depending only on $c_{4}$, $s,t,u$ and
$\ell \cdot k_{4}$.

Analytically evaluating it with \texttt{Mathematica}, we obtain a
remarkable simplification of the determinant which reduces to a
single term
\begin{equation}
  \label{eq:jacsolved}
  J=-64 q i \pi^7  t^2 (\ell\cdot k_4)^2\,.
\end{equation}
Replacing $q$ \eqref{eq:scalqpk2pk3} as well as $c_2$ and $c_3$, we obtain
\begin{equation}
  \label{eq:jacfinal}
{J}=
-{16 i \pi^3} 
\frac{\ell^2 (\ell\cdot k_2) (\ell\cdot k_3)}{t}
{(\pi u-i \ell\cdot k_4)(\pi s+i \ell\cdot k_4)}\,.
\end{equation}

At this point, we see already an interesting
combination that appears at the right end of the last expression.
This is highly reminiscent of a combination of two IR boxes in
fig.~\ref{fig:boxes}, up to a re-normalisation
of $\ell\to2i \pi \ell$.

The last step of the prescription is to sum over the solutions of the
scattering equations. At four-point, we already mentioned that two
solutions contribute to this IR limit, the one we described and
its complex conjugate. Hence we need to sum the inverse
Jacobian and its value for the complex conjugate solution. 
To do this, a last subtlety has to be addressed. The Jacobian contains a
$\partial_{\tau}$ derivative, which is not a holomorphic operation on
$q$. Therefore, the evaluation of the Jacobian for the second
solution, which we denote $\tilde J$, is obtained by exchanging the
$z_{i}$'s and $q$ for their complex conjugate, while not complex
conjugating the $i$ originating from
$\partial_{\tau}=2i\pi q \partial_{q}$. The final result is
\begin{equation}
  \label{eq:jacsum}
  \frac1{J}+  \frac1{\tilde{J}}=
\frac{-1}{(16 i \pi^3)\ell^2 (\ell\cdot k_2) (\ell\cdot k_3)}
\frac{2\pi^2stu+2(\ell.k_4)^2}
{((\pi u)^2+ (\ell\cdot k_4)^2)((\pi s)^2+( \ell\cdot k_4)^2)}\,,
\end{equation}
which is exactly the sum of symmetrized boxes \eqref{eq:totIR},
after taking $\ell \to 2 i \pi \ell$.
Note that nowhere in this computation the spacetime dimension was used explicitly. This contributes to make us believe that it is
actually independent of it, and that the integral eq.~\eqref{eq:I4} is
actually well defined in any dimension. 

\subsection{Extension to $n$ points}
\label{sec:extension-n-point}
Remarkably, the solution presented in the previous section extends
straightforwardly to $n$ points, at least qualitatively.
Going again to the limit where three adjacent propagators go on shell,
we use the ansatz of eq.~\eqref{eq:scalingpkpk23}.

The qualitative behaviour follows from the fact that the arguments for
factoring $q$ out of the Jacobian still hold, and so does the scaling
obtained in eq.~(\ref{eq:scalqpk2pk3}). Therefore, we have immediately
that \textit{the Jacobian possess terms with the qualitative IR
  behaviour expected of scalar $n$-gons.} This streghtens the
interpretation of the scalar integrals of the type of
eq.~(\ref{eq:I4}) as scalar $n$-gons, that can be defined in any dimension.

It is even possible to actually extract information on the form of
$z_{2}$ and $z_{3}$. The scattering equations for $z_{2}$ and $z_{3}$
are solved exactly in the same manner as they were in
eq.~\eqref{eq:sceqns23IR}, more precisely they read
\begin{equation}
\label{eq:sceqns2IRnpt}
\begin{aligned}
  2\ell\cdot k_2 - i k_{1}\cdot k_{2} (1+2\ell \cdot k_2 c_2)
- &i k_{2}\cdot k_{3} (1+2\ell \cdot k_2\ell \cdot k_3 c_2c_3)\\
- &i \sum_{j=4}^{n} k_{2}\cdot k_{j} (1+2\ell \cdot k_2 c_2c_j)=0\,,
\end{aligned}
\end{equation}
and
\begin{equation}
\label{eq:sceqns3IRnpt}
\begin{aligned}
  2\ell\cdot k_3 - i k_{1}\cdot k_{3} (1+2\ell \cdot k_3 c_3)
+ &i k_{2}\cdot k_{3} (1+2\ell \cdot k_2\ell \cdot k_3 c_2c_3)\\
+ &i \sum_{j=4}^{n} k_{3}\cdot k_{j} (1+2\ell \cdot k_3 c_3c_j)=0\,,
\end{aligned}
\end{equation}
where the $c_{j}$ for $j\geq4$ are defined just like $c_{4}$ in
eq.~(\ref{eq:c4def}).
These equations can be solved as in eq.~(\ref{eq:c2c3}), using
momentum conservation and replacing $k_{2/3}\cdot k_{4 }c_{4}$ by the
sum $\sum_{j=1} k_{2/3}\cdot k_{j} c_{j}$. The unknowns $c_{2}$ and
$c_{3}$ being expressed in terms of $c_{4}$ as
\begin{equation}
  \label{eq:c2c3-c4}
    c_{2}=\frac{1}{i\pi(k_{1}\cdot k_{2}+ k_{2}\cdot k_{4} c_{4})}\,,\qquad
    c_{3}=\frac{-c_{4}}{i\pi(k_{1}\cdot k_{2}+ k_{2}\cdot k_{4} c_{4})}\,,
\end{equation}
and it is now straightforward to replace $c_{4}$ by its $n$-point value.

A more precise statement would require solving for the remaining
$c_{j}$, which quickly becomes difficult for high values of $n$.

\section{Relation to Gross \& Mende}
\label{sec:discussion}

In this final section we wish to discuss the results of the previous
section and in particular explain that the one-loop ambitwistor saddle
point is the same as the Gross-Mende saddle point, though modified by
reintroducing in the string amplitude an explicit loop momentum zero
mode integral. This also gives a cross check of the validity of the
change of normalisation required by the previous computation.

\subsection{Changing the normalisation}
\label{sec:chang-norm}

Before discussing the $2i\pi \ell$ normalisation, let us first observe
that the scattering equations possess a
$z,q\leftrightarrow \bar z,\bar q$ symmetry when written in terms of
the holomorphic propagator, as a consequence of their holomorphy.
This elucidates the reason why we found two solutions complex
conjugate to one another in the previous section. This symmetry
should hold at all loop order and any number of points, and probably
induce, strictly speaking, a factor of $1/2$ in the lower bound on the
number of solutions at one loop in eq.~(\ref{eq:nsols}).

Adopting the $2i\pi \ell$ normalisation has the obvious consequence
that when conjugating the equations, the loop momentum flip
signs. This means that given a set of solutions for loop momenta
$\ell$, new solutions can be obtained simply by flipping the sign of
$\ell$. These correspond to the same configuration but with the loop
momentum flowing in the opposite direction in the loop. 

Another consequence of adopting this normalisation is that now, the
solutions are \textit{purely imaginary}, which is good since in the
end we want to be sure that the integrand will be real. Summing over
the solution and its complex conjugate is admittedly good enough for
this, but solutions lying on a line have some other advantages, in
particular for numerical purposes.

\subsection{Gross \& Mende limit and the electrostatics analogy}
\label{sec:electr-analogy}
\begin{figure}[t]
  \centering
  \begin{tikzpicture}
    \coordinate (Origin)   at (0,0);
    \coordinate (XAxisMin) at (-1,0);
    \coordinate (XAxisMax) at (3.2,0);
    \coordinate (YAxisMin) at (0,-0.8);
    \coordinate (YAxisMax) at (0,3.4);
    \draw [thick,-latex] (XAxisMin) -- (XAxisMax);
    \draw [thick,-latex] (YAxisMin) -- (YAxisMax);
    \pgftransformcm{1}{0.}{0.3}{1.2}{\pgfpoint{0cm}{0cm}}
    \draw[style=help lines,dashed] (-1,-0.5) grid[step=2cm] (2.7,2.7);
    \foreach \x in {0,1}{
      \foreach \y in {0,1}{
        \node[draw,circle,inner sep=2pt,fill] at (2*\x,2*\y) {};}}
    \foreach \xx in {1/2}{
      \foreach \yy in {0,1}{
        \node[draw,circle,inner sep=2pt] at (2*\xx,2*\yy) {};}}
    \foreach \x in {0,1}{
      \foreach \y in {1/2}{
        \node[draw,rectangle,inner sep=2pt] at (2*\x,2*\y) {};}}
    \foreach \x in {1/2}{
      \foreach \y in {1/2}{
        \node[draw,diamond,inner sep=1.3pt] at (2*\x,2*\y) {};}}
  \end{tikzpicture}
  \caption{Gross \& Mende equilibrium; the charges should be
  placed at half-periods of the lattice.}
  \label{fig:GM}
\end{figure}
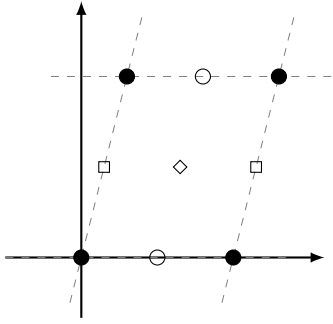

Let us now come to the relationship between the ambitwistor and the
Gross \& Mende saddle point.  Gross and Mende studied the high energy
limit of closed string amplitudes. The type II 4-graviton amplitude
in 10 dimensions reads
\begin{equation}
  \label{eq:4gravtypeII}
  \int_{\mathcal{F}}\frac{\d^{2}\tau}{\Im \tau^{2}} \int
  \prod_{i=2}^{4}\frac{\d^{2} z_{i}}{\Im \tau} 
\left|e^{2
    \alpha'\sum_{i,j}k_{i}\cdot k_{j} G_{ij}}\right|^{2}
\end{equation}
The saddle points of this integral when $\alpha'\to\infty$ are
obtained when the energy
$\mathcal{E}=\alpha'\sum_{i,j}k_{i}\cdot k_{j} G_{ij}$ is an extremal
with respect to variations of the moduli $z_{i}$ and $\tau$ ($G_{ij}$
is defined to be $G(z_{ij},\tau)$). The leading contribution was
claimed to come from the saddle corresponding to most symmetric way to
arrange the four charges on the torus; these should sit at
half-periods of the lattice, such that
$\{z_{1},z_{2},z_{3},z_{4}\}=\{1/2,\tau/2,(\tau+1)/2,0\}$, (up to
permutations), as pictured in fig.\ref{fig:GM}.

With this choice, it was explained by Gross \& Mende that not only the
$\partial_{z}\mathcal{E}$ scattering equation vanish, but every single
term of the sum is actually a zero of the propagator, hence vanish as
well. The last saddle point equation, $\partial_{\tau}\mathcal{E}$ is
solved by the following condition
\begin{equation}
  \label{eq:qGM}
  \frac{\theta_{2}(0,\tau)^{4}}{\theta_{3}(0,\tau)^{4}}=-\frac{u}{s}\,.
\end{equation}

One would like to think of these saddle point equations as the
$P\cdot k$ and $P^{2}$ scattering equations, respectively. However a
crucial ingredient is missing; there is no loop momentum. This
actually can be cured by reverse engineering a string amplitude with
explicit loop momentum, see for instance the classical
ref.~\cite{D'Hoker:1988ta}. Starting from \eqref{eq:4gravtypeII}, one
has to undo the $\partial X$ zero mode integral and write
\begin{equation}
  \label{eq:4gravtypeIIzeromode}
  \int\frac{\d^{10}\ell}{(2\pi)^{10}}
  \int_{\mathcal{F}}\frac{\d^{2}\tau}{(\Im \tau)^{2-5}} \int
  \prod_{i=2}^{4}\frac{\d^{2} z_{i}}{\Im \tau} 
\left|e^{i \pi \tau \ell^{2} + 2 i \pi \sum_{i=1}^{4}\ell\cdot k_{i} z_{i}}\right|^{2}
\left|e^{2
    \alpha'\sum_{i,j}k_{i}\cdot k_{j} S_{ij}}\right|^{2}\,,
\end{equation}
where the $-5$ in the exponent of $\Im\tau$ comes from the
reintroduction of the loop momentum gaussian integral. It is easily
checked on this expression that integrating out back again the zero
mode part provides the expected non-holomorphic correction to the
propagator.

This provides an alternative energy $\tilde{\mathcal{E}}(\ell)$ which
explicitly depends on the loop momentum. 
Analysing now the saddle point of this amplitude, we have two options;
either we ask for a saddle point in the $\ell$ direction, i.e. we
add the $\partial_{\ell}\tilde{\mathcal{E}}(\ell)=0$ equation, or we
leave unfixed the integration over the loop momentum and solve the
saddle point at each values of $\ell^{\mu}$. The former gives
\begin{equation}
  \label{eq:saddleloop}
  \ell_{*}^{\mu}= \sum_{i=1}^{n}k_{i}^{\mu} \frac{\Im z_{i}}{\Im \tau}
\end{equation}
which, once inserted in the
$\partial_{z/\tau}\tilde{\mathcal{E}}(\ell)=0$ saddle conditions,
consistently gives back the Gross \& Mende saddle point equations. The
other latter option yields the one loop scattering equations proposed
by \cite{Adamo:2013tsa}, whose IR analysis was the focus of this
work.

A striking difference between the two approaches is that the value of
$\tau$ is simply fixed by the external kinematics. So far, we have not
succeeded in understanding how the Gross \& Mende saddle point should
be deformed in the presence of a loop momentum and holomorphic
propagators, nor have we understood the physical relevance of the
existence of a preferred value at a threshold for the loop momentum.
\begin{equation}
  \label{eq:lGM}
  \ell_{*}^{\mu}=k_{2}^{\mu}+k_{3}^{\mu}
\end{equation}
We can simply obervse that modular invariance acts by permuting which
scattered particles sit on the half periods, thereby changing the
previous loop momentum to a different threshold.

It is also an amusing question to wonder what kind of electrostatic
problem the ambitwistor saddle point corresponds to, as we now possess
an energy $\tilde{\mathcal{E}}(\ell)$ which is extremized by the
scattering equations. In particular, one would want to understand what
creates the contribution of the loop momentum in the equations.
The energy is still invariant when winding around the $A$- and
$B$-cycles, due to momentum conservation which corresponds to a charge
neutrality condition. 
However, the individual interactions between the charges themselves
are not invariant anymore when one winds around the $B$ cycle of the
torus, $z_{i}\to z_{i}+\tau$, as the $\tau$ periodicity was guaranteed
by the non-holomorphic part of the propagator. A plausible
electrostatic problem which should have such a kind of solution may be
to consider two infinitely long wires with lineic charges
$\pm\ell/\Im \tau$ located at $z=0$ and $z=1/2$, in the lattice
$\mathbb{C}/(\mathbb{Z}+\tau\mathbb{Z})$. These create a constant
potential proportional to $\ell$ in the region $0<z<1/2$, and 0 outside.

\section{Outlook}
\label{sec:outlook}

The scattering equations are at the core of tree-level scattering of
massless particles. In their D dimensional guise they were introduced
and studied in the series of works
\cite{Cachazo:2013gna,Cachazo:2013hca,Cachazo:2013iaa} where formulas
for scalars, gluons and gravitons were obtained, collectively called
CHY formulas, and more recently for Einstein-Yang-Mills in
\cite{Cachazo:2014nsa}. Since then the tree-level scattering equations
have been studied in several contexts. A proof of the equivalence of
the CHY formulas to the scattering amplitudes was given in
\cite{Dolan:2013isa}, a polynomial form for the scattering equations
which makes transparent their number of solutions and an algorithm to
compute them was given in \cite{Dolan:2014ega}. Also, a different
appraoch to solving the one-loop scattering equations was proposed in
\cite{cachazo2014}.

The relationships between the scattering equations and
colour-kinematics duality \cite{Bern:2008qj,Bern:2010ue} was explored
in the original works and was further explored in in
\cite{Monteiro:2013rya,Naculich:2014rta,Naculich:2014naa}, its
relation to string theory amplitudes was studied in
\cite{Yuan:2014gva,Bjerrum-Bohr:2014qwa}. The scattering equations as
well as the CHY formula were explained to originate from the
ambitwistor string introduced in \cite{Mason:2013sva}. In this
approach the scattering equations appears from the usual BRST gauge
fixing procedure, this allowed generalisations of the scattering
equations to curved spacetimes in \cite{Adamo:2014wea} and, crucial
for the present work, to loop level in \cite{Adamo:2013tsa}. The
scattering equations were already discussed in their four dimensional
guise in the context of the original twistor string in
\cite{Witten:2004cp} where its geometrical meaning was already known.

In the context of string theory, the scattering equations go back to
the beggining of the subject in the work of Fairlie and Roberts
\cite{Fairlie:1972}, most notably they appear in the high energy limit
of string scattering where they also localise the string integrals
through the steepest descent approximation \cite{Gross:1987ar}.  Also
recently they were used in \cite{Dvali:2014ila} in the context of high
energy gravitational scattering, where a scenario, called
``classicalization'' \cite{Dvali:2010jz}, different than the usual
string exponentially soft behaviour is used to regulate the non
unitarity of the process.  In a slightly indirect way, we may also
note that one-loop maximally supersymmetric Yang-Mills amplitudes are
known explicitly in the pure spinor formalism \cite{Mafra:2012kh} (see
also recent progress in
\cite{Mafra:2014gja,Mafra:2014gsa,Mafra:2014oia}) and are expressed in
terms of a basis of tree-level building blocks
\cite{Mafra:2011nv,Mafra:2011nw,Mafra:2011kj}. Since the latter are
known to arise also from the CHY formalism, this gives an indirect way
to implement a scattering equation prescription at the one-loop level.

Let us conclude this work with a short outlook.
Firstly, the question to determine the number of solutions is still
open after this work. Any argument in this direction is of crucial
importance. Secondly, the UV behaviour of the solutions we
investigated here should be analysed; a computation similar to the one
presented here, if doable, should bring a UV asymptotics of the form
$1/\ell^{2n}$ at $n$ points for the Jacobian.\footnote{We of course expect the
Pfaffians of the supergravity amplitude to give additional factors of
$\ell^{2n-8}$ to produce the correct total UV behaviour of maximal
supergravity amplitudes}
We already commented in several instances in the text why we believe
that integrals of the form of eq.~(\ref{eq:I4}) defined for $n$ points
correspond to scalar $n$-gons;
\begin{itemize}
\item First, we saw in sec.~\ref{sec:computation-jacobian} that the
  four point integral seems not to depend so much on the spacetime dimension,
\item Second, the IR behaviour also does not depend on the number of
  external particles, as explained in sec.~\ref{sec:extension-n-point}.
\end{itemize}
This gives hope that this formalism will apply very generally to all
sort of amplitudes.
In particular, an extension of the solution presented in this text to
heterotic ambitwistor models would be very interesting. The difficulty
in doing so will be to isolate the Yang-Mills degrees of freedom
running in the loop.

\paragraph{Acknowledgements}

We would like to acknowledge the contribution of Yvonne Geyer for her
collaboration at an initial stage of this project.  
We would also like to thank Arnab Rudra for interesting discussions on
the electrostatic analogy, and Tim Adamo, Pierre Vanhove and
especially David Skinner and for many enlightening discussions,
interest and support and various comments on the text on this project.

We would finally like to thank, both for financial support and
organisation, the organisers of the Stony Brook 2013 workshop ``The
Geometry and Physics of Scattering Amplitudes'', where this project
was initiated.

This work is supported by the ERC Grant 247252 STRING. The work of EC is supported by the Cambridge Commonwealth, European and International Trust.

\appendix

\bibliography{./bibliography.bib}
\bibliographystyle{JHEP}

\end{document}